# Analysis of homogeneity of 2D electron gas at decreasing of electron density

A. A. Sherstobitov, G. M. Minkov, A. V. Germanenko*, O. E. Rut*, I.V. Soldatov*, B.N. Zvonkov**

Institute of Metal Physics RAS, Ekaterinburg, Russia
* Ural State University, Ekaterinburg, Russia
** Physical-Technical Research Institute, University of Nizhni Novgorod, Nizhni Novgorod, Russia

We investigate the gate voltage dependence of capacitance of a system gate – 2D electron gas (C-$V_g$). The abrupt drop of capacitance at decreasing concentration was found. The possible reasons of this drop, namely inhomogeity of electron density distribution and serial resistance of 2D electron gas are discussed. Simultaneous analysis of gate voltage dependences of capacitance and resistance has shown that in heavily doped 2D systems the main role in the drop of capacitance at decreasing concentration plays the resistance of 2D gas. It is found that the investigated systems remains homogeneous down to the low temperature conductivity about $(10^{-2}-10^{-3})e^2/h$.

## 1. Introduction

The investigation of quantum effects in 2D systems has a great advantage due to the ability to control the concentration of charge carriers continuously by varying voltage on the gate electrode. With this variation, the role of electron-electron interaction changes as well as the role of disorder. The contribution from this interaction is determined by the parameter $r_s$ – the ratio of the coulomb interaction energy at an average distance $E_C=e^2n^{1/2}$ to the Fermi energy $E_F$. The role of disorder is governed by the parameters $h/\tau$ and $U$, where $\tau$ is the momentum relaxation time and $U$ is the characteristic energy scale of smooth random disorder.

In all cases with a decreasing electron density the mobility decrease too. There are several different physical mechanisms of such behavior. When $E_C>h/\tau$, $U$, the decreasing of carriers concentration can lead to the Wigner crystallization. If , $h/\tau> E_C,U$, the Anderson localization plays the central role, and at low temperatures, one can get hopping conductivity over the states localized due to this effect. Finally, when the fluctuations of smooth random potential are significant ($eU>h/\tau$, $E_C$), the decrease of electron density may cause the formation of droplet structures – "metallic" regions separated by barriers – and low temperature conductivity will be determined by carrier tunneling through those barriers.

Usually, a decrease in the electron density causes all three parameters to change, thus, in order to appropriately interpret experimental data, one should know which of those mechanisms plays the main role. For example, in [1], the authors interpret the abrupt drop of conductivity σ (for σ <=10$G_0$, $G_0=e^2/\pi h$) with decreasing electron density in terms of a "droplet structures" model. At the same time, in [2-5] the electron gas is believed to be homogeneous down to much lower values of conductivity (σ=$(10^{-2}- 10^{-3})G_0$), and the strong temperature dependence was explained in terms of Anderson localization. So one need a method of analysis of homogeneity of electron gas to adequately interpret our data.

In the present paper, we investigate C-$V_g$ dependences on structures with large disorder (with delta-doped quantum well) at different frequencies and temperatures. The data was adequately described assumption of homogeneity of the electron gas up to conductivity values of $(10^{-3}\ 10^{-2})\ G_0$.

## 2. Experimental details

The investigated heterostructures were grown by metalorganic vapor-phase epitaxy on a semi-insulating GaAs substrate and consist of 0.5-mkm-thick undoped GaAs epilayer, a 8nm In$_{0.2}$Ga$_{0.8}$As quantum well with Sn δ layer situated in the well center, and a 200nm cap layer of undoped GaAs. Several samples were mesa-etched from each wafer into standard Hall bars and then an Al gate electrode was deposited onto the cap layer by thermal evaporation. The Hall bar geometry and shape of the Al-gate electrode are presented in the inset of fig.1. The electron density n and mobility $\mu$ at $V_g=0$ where in the range $(1-2) \times 10^{12} cm^{-2}$ and $(1500-2500) cm^2/Vs$ respectively. The investigations of the transport properties of these structures can be found in [5]. DC conductivity (4-probe method: current passed from contact 1 to 2, voltage drop was measured between 3-4) was measured in the linear regime, the range of which was derived from a study of I-V dependences. The gate was deposited so that the potential contacts remained uncovered. Capacitance and DC conductivity measurements were carried out during the same cooldown.

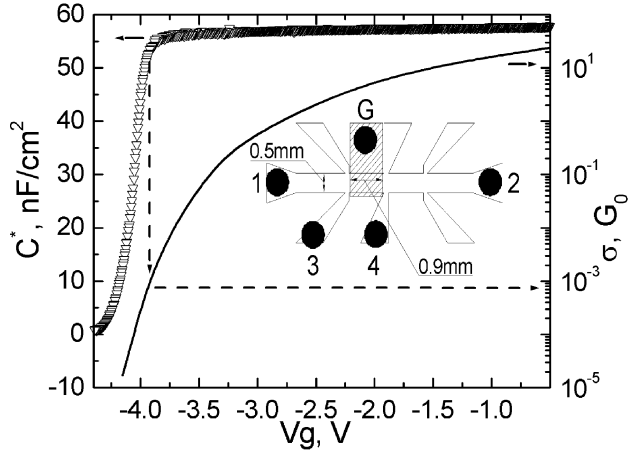

*Fig. 1: The gate voltage dependence of the effective capacitance C* in configuration 1-G. Inset: The Hall bar geometry, contacts and gate.*

To measure the impedance Z of the sample we use digital lock-in amplifier. To obtain the impedance dependence on gate voltage we apply the sum of AC and DC voltages ($V_{AC}$ and $V_g$) to the sample and measure the AC current ($I_{AC}$) as a function of $V_g$. The complex impedance was calculated as $Z=V_{AC}/I_{AC}$, taking in account the phases of I and V. To make sure that the sample remains macroscopically homogeneous (particularly at low carrier density) three configurations of sample connections were used during the experiment: the measurement of C-$V_g$ was carried out between the gate and contacts 1 or 2, or shorted out 1 and 2, the configurations 1-G, 2-G and 1,2-G respectively (see inset in fig. 1). The amplitude of the AC signal was less than 15mV. The increasing of AC amplitude leads to the distortion of the shape of gate voltage dependence. The effective capacitance C* was calculated using the following expression: C*=1/(w*Im(Z)). We note the C* - effective capacitance because this is not capacitance of the sample but the capacitance in equivalent circuit consisting of capacitance in series with resistance. Really the equivalent circuit of the sample is more complicated. For the analysis we will compare the same parameter C* for sample and model.

### 3. Results and discussion

Several different samples were investigated. All the samples show the similar results. Here we discuss one of them, number 4261 with the zero gate voltage electron density $n=1.8 \times 10^{12} cm^{-2}$ and mobility $\mu=1700 cm^2/Vs$. The gate voltage dependence of specific (per unit area) effective capacitance C* at a temperature of liquid helium 4.2K is presented in fig.1. It is seen that in the range of $V_g$ from 0 to -3.5V, the capacitance depends almost not at all on bias[1]. With further lowering of Vg, at value of -3.6 V one can observe the abrupt drop of the capacitance.

It should be mentioned that in the range of Vg (0 - -3.3)V the experimental value of capacitance C* is close to the geometrical value of $C_g = \frac{\varepsilon \varepsilon_0}{d}$, where the dielectric constant of GaAs is $\varepsilon=12.6$, d=200nm, is the thickness of the cap layer. This is in good agreement with the gate voltage dependence of carrier density
$dn/dV_g = C_g/e \approx 3.5 \cdot 10^{11} 1/(cm^2 V)$ .

A very similar C-vs-Vg dependence was obtained in [6], where the drop of C was interpreted as a consequence of formation of dielectric regions in the 2D gas (droplet structure formation). Before analyzing the role of this mechanism, let us take in to account the gate voltage dependence of DC-conductivity (fig. 1). One can see that a noticeable decrease of C* takes place when the conductivity of 2D gas becomes smaller than $(10^{-2}-10^{-3})G_0$. It is clear that with an increase in 2D gas resistivity, the imaginary part of Z depends not only on the 2D gas - gate capacitance, but also on the value of serial in-plane resistance. In this case, the equivalent circuit could be presented as a distributed parameters line, depicted in the inset in figure 2. In homogeneous case all elements are equal. The impedance of such a circuit can be written as:

$$Z(\omega, r, c, L) = \sqrt{\frac{r}{i\omega c}} \coth(L\sqrt{i\omega cr}) \quad (1)$$

---

[1] Actually in this gate voltage range there is decrease of about 2 percent in the capacitance. Such a dependence was observed in different structures. In our opinion, the small decrease in the capacitance is not due to the compressibility of the 2d gas. Such a dependence may be a consequence of deep level recharging.

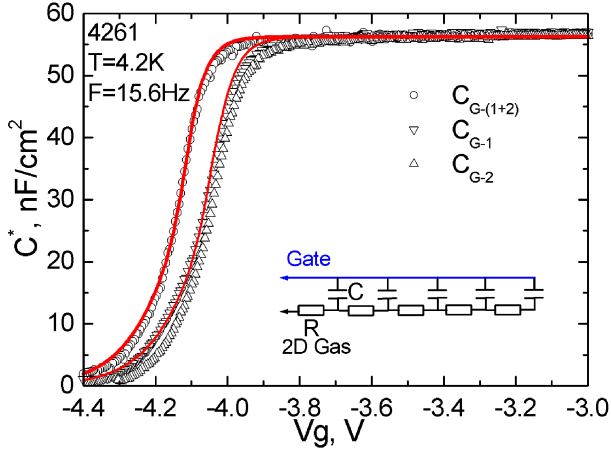
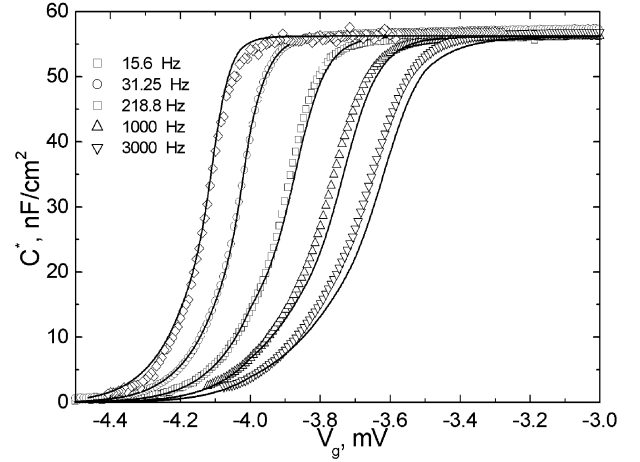

*Fig. 2: The gate voltage dependencies of C*, for all three configurations (see text for details). Solid lines (color online) show the calculated dependencies (eq. 1). Inset: The equivalent circuit for the structures investigated. In homogeneous case all the elements R and C are equal.*

*Fig. 3: The gate voltage dependence of the effective capacitance, measured at different frequencies. Solid lines show the calculated dependencies (eq. 1).*

Here, $c=C/L$ and $r=R/L$ are the capacitance and resistance per unit length, C and R are the geometrical capacitance and DC resistance, $\omega$ is the circular frequency, L is the length of the gate electrode. In the case of short-circuited contacts 1 and 2, the line effectively becomes twice as short and twice as wide, and one should use $c=2C/L$, $r=R/2L$, $L/2$. To compare the predictions of this model with experiment, we calculated the $C^*(V_g)$ dependence using the measured $R(V_g)$ dependence. The resulting curves for all configurations (1-G or 2-G and 1,2-G) are shown in fig. 1. It is seen that experimental curves in configurations 1-G and 2-G are close together. This fact prove the absents of strong inhomogeneities in the sample. We can see that this simple model works well in all configurations. It should be noted, that this model works well down to conductivity values about $(10^{-3} – 10^{-4}) G_0$. (see fig.1)

The fact that the drop in capacitance is caused by the effect of the serial resistance of the 2D gas can be seen more clearly from the following speculations. The gate dependence of Z is governed by the relation between R and $1/(\omega C)$. So the drop in capacitance should move to more-negative gate voltages with increasing temperature (due to temperature dependence of conductivity) and to less-negative values of voltage with increasing frequency.

The experimental dependencies and theoretical curves calculated by equation (1), are shown on the figs 3,4 for different temperatures and frequencies. It can be seen that in both cases there is good agreement between the data and theoretical curves, calculated with the assumption of homogeneity of electron gas.

Another proof of the validity of the distributed parameters line model can be found in the following way. Figure 5 presents the $C^*(\sigma)$ dependences obtained at different temperatures (when conductivities at certain $V_g$ were substantially different: up to 2 orders of magnitude at $V_g=-4V$). It can be seen (fig. 5) that at this scale all experimental dependences fall on one curve and are in agreement with theoretical calculations.

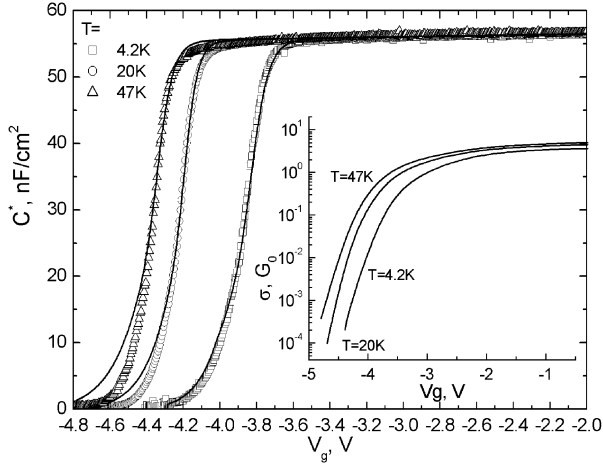
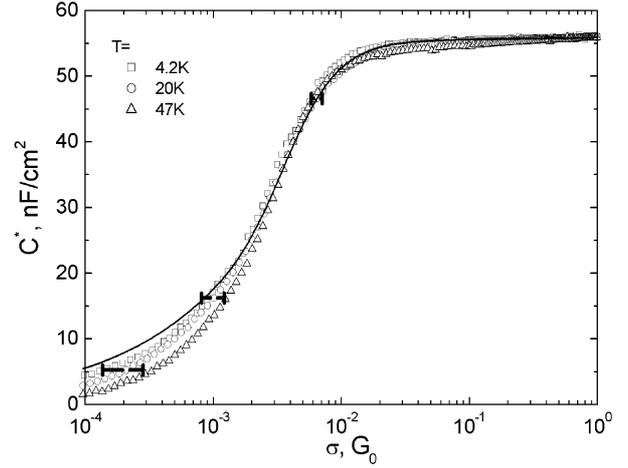

Fig. 4: The gate voltage dependence of the effective capacitance, measured at different temperatures. Inset: The gate voltage dependence of the DC conductivity at different temperatures.

Fig. 5: The conductivity dependence of effective capacitance C*, measured at different temperatures.

### 4. Conclusion

In conclusion, this analysis shows that in the investigated structure the 2D gas remains homogeneous, at least until a conductivity of $10^{-3}$ $G_0$, and that the drop in C* takes place due to the increase of 2D gas resistivity instead of the formation of dielectric regions. It should be mentioned that such behavior (the fact that the 2D gas remains homogeneous while the value of conductivity is above $10^{-2} G_0$) is a consequence of the sample's features: high electron density and high scattering rate occur because of impurities which have been placed directly into the quantum well. In structures with remote impurities, the very similar behavior of C* with decreasing electron density [6] is due to the formation of droplet structures. It becomes obvious, if one take a look at the value of conductivity, at which the drop of $1/(\omega Im(Z))$ occurs. In [6] it happens at conductivity (1-10)$G_0$, when the resistivity of 2D gas has no influence on C*, according to the distributed parameters theory (eq. 1). Thus simultaneous analysis of $C(V_g)$ and $R(V_g)$ dependencies allow us to test the homogeneity of 2D electron gas.

### 5. Acknowledgment

The authors are grateful to Paul Boley for assistance in translation. The work is supported by RFBR grants 08-02-00662, 08-02-91962, 09-02-789, 09-02-12206, 10-02-00481, 10-02-91336.

**References.**

[1] S. Das Sarma, M. P. Lilly, E. H. Hwang, L. N. Pfeiffer, K. W. West, and J. L. Reno, Two-Dimensional Metal-Insulator Transition as a Percolation Transition in a High-Mobility Electron System, Phys. Rev. Lett. **94** (2005) 136401.

[2] G. M. Minkov, O. E. Rut, A. V. Germanenko, A. A. Sherstobitov, B. N. Zvonkov, E. A. Uskova, and A. A. Birukov Quantum corrections to conductivity: From weak to strong localization Phys. Rev. B **65** (2002) 235322.

[3] G. M. Minkov, O. E. Rut, A. V. Germanenko, A. A. Sherstobitov, V. I. Shashkin, O. I. Khrykin, and B. N. Zvonkov, Electron-electron interaction with decreasing conductance, Phys. Rev. B **67** (2003) 205306.

[4] G. M. Minkov, A. V. Germanenko, and I. V. Gornyi, Magnetoresistance and dephasing in a two-dimensional electron gas at intermediate conductances, Phys. Rev. B **70** (2004) 245423

[5] G. M. Minkov, A. V. Germanenko, O. E. Rut, A. A. Sherstobitov, and B. N. Zvonkov, Giant suppression of the Drude conductivity due to quantum interference in the disordered two-dimensional system GaAs∕InxGa1−xAs∕GaAs, Phys. Rev. B **75** (2007) 235316

[6] G. Allison, E. A. Galaktionov, A. K. Savchenko, S. S. Safonov, M. M. Fogler, M.Y. Simmons, and D. A. Ritchie Phys. Rev. Lett. **96** (2006) 216407.